# A Simulation of Optimal Dryness When Moving in the Rain or Snow Using MATLAB

Neil Zhao[a,*], Emilee Brockner[a], Asia Winslow[a], Megan Seraydarian[a]

[a]*Thomas Jefferson University, Philadelphia, Pennsylvania, USA*

The classic question of whether one should walk or run in the rain to remain the least wet has inspired a myriad of solutions ranging from physically performing test runs in raining conditions to mathematically modeling human movement through rain. This manuscript approaches the classical problem by simulating movement through rainfall using MATLAB. Our simulation was generalizable to include snowfall as well. An increase in walking speed resulted in a corresponding decrease in raindrop and snowflake collisions. When raindrops or snowflakes were given a horizontal movement vector due to wind, a local minimum in collisions was achieved when moving in parallel with the same horizontal speed as the raindrop; no local minimum was detected with antiparallel movement. In general, our simulation revealed that the faster one moves, the drier one remains.

---

*E-mail of corresponding author: neil.zhao@students.jefferson.edu

# Introduction

Should one walk or run in the rain to stay the least wet? This question has been a curiosity since at least the 1970s [1]. The popular television program *Mythbusters* featured this problem in one of its earliest episodes and came to the conclusion that one should walk rather than run [2]. However, other approaches to this question using mathematical approximations of human movement through rain [3–5] have concluded that the best strategy is to run as quickly as possible toward your destination. Further studies into this topic have generally indicated that moving at higher speeds is the proper choice for staying as dry as possible [6].

We have approached this problem by constructing a MATLAB based simulation of movement through rain. We have also expanded the scope of our simulation to include snowfall. MATLAB is a mathematical computational program that is commonly used in academic and industrial environments. As such, it plays a crucial role in enlarging the computational arsenal beyond analytical solutions to include numerical solutions. The use of MATLAB in approaching this curious problem can therefore be an educational tool to demonstrate the power of computational software in answering questions that would otherwise require extensive mathematical rigor.

Through our simulation, we demonstrated that overall moving at higher speeds resulted in fewer collisions with raindrops. This also applied to collisions with snowflakes when the simulations conditions were adjusted to model snowfall.

# Methods

The simulation of precipitation was created using MATLAB R2019a (MathWorks, Natick, MA). Precipitation was modeled over a 1 X 1 cartesian grid, with a moving rectangular shape used to approximate a human form. Variables that were allowed to be adjusted included the number of raindrops/snowflakes in the field

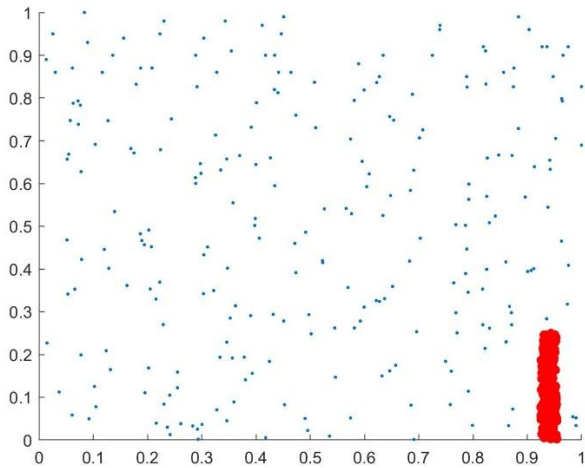

A frame of the simulation

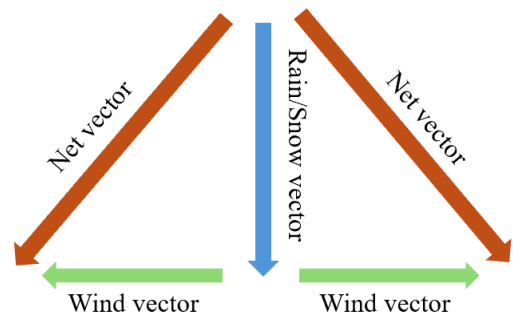

Schematic of rain/snow and wind vectors

per frame, speed of the human approximation, horizontal wind speed, vertical rain speed, vertical snowfall speed, angle with respect to the vertical for rain and snowfall. The human approximation was composed of 200 randomly generated points inside [0,1] x [0,1], which was then compressed along the x-axis to 0.03 the original length and along the y-axis to 0.25 the original length. Every simulation began with the human approximation at x=1 and moving toward x=0. At every frame, the distance between each of the 200 points that composed the human approximation and each raindrop/snowflake was calculated. Any distance less than or equal to 0.01 was considered a collision. Any raindrop/snowflake subject to a collision was removed from the simulation and reset at a random location along y=1. Any raindrop/snowflake that reached y=0 without a collision was reset at a random location along y=1. Any raindrop/snowflake that reached x=0 without a collision was reset to a random location along x=1. Any raindrop/snowflake that reached x=1 without a collision was reset to a random location along x=0. An option was given to allow oscillatory behavior of the precipitation, but this was not utilized during any segment of the simulation. The total number of collisions for each complete run of the human approximation between x=1 and x=0 was recorded. Each speed of the human approximation was repeated 10 times and averaged.

Complete MATLAB code for a run with wind speed set to 1/10X of rain speed is provided below with annotations.

```matlab
function rainrun(drops,angle,trials,v)

%drops is the number of raindrops
%angle is the angle of rain/snow
%trials is the total factor increase in speed of person by increments of 1
%0.1*v is the speed of person

repeats=10;
pop=200;
runs=ceil(1/v);
wind=-0.001;
rainspeed=0.01;

rainx=rand(1,drops);
rainy=rand(1,drops);
rain=zeros(2*runs,drops);
angles=(rand(1,drops)*angle*2-angle)*pi/180;

xdis=zeros(pop,drops);
ydis=zeros(pop,drops);

hits=zeros(trials,runs,repeats);

for k=1:repeats
    for j=1:trials
        me=zeros(pop,2,runs);
        me(:,:,1)=rand(pop,2);
        me(:,1,1)=0.03*me(:,1,1)+1;me(:,2,1)=0.25*me(:,2,1);
        i=1;
        while mean(me(:,1,i))>=0
            me(:,:,i+1)=me(:,:,i)-j*0.1*v*repmat([1 0],pop,1);

            rain(1,:)=rainx;rain(2,:)=rainy;

            %step is for oscillating the rain drops
            %step is multiplied after rainspeed below
            %step=rand(1,drops);step(step<0.5)=1;step(step>=0.5)=1;
            rain(i*2+1,:)=rain(i*2-1,:)+rainspeed*tan(angles)+wind;
            rain(i*2+2,:)=rain(i*2,:)-rainspeed;
            a=find(rain(i*2+2,:)<=0);
            e1=find(rain(i*2+1,:)<=0);e2=find(rain(i*2+1,:)>=1);
            rain(i*2+1,a)=rand(1,length(a));rain(i*2+2,a)=ones(1,length(a));
            rain(i*2+1,e1)=ones(1,length(e1));rain(i*2+2,e1)=rand(1,length(e1));
            rain(i*2+1,e2)=zeros(1,length(e2));rain(i*2+2,e2)=rand(1,length(e2));
            angles(1,[a e1 e2])=(rand(1,length([a e1 e2]))*angle*2-angle)*pi/180;

            xdis(:,:)=rain(i*2-1,:)-me(:,1,i);ydis(:,:)=rain(i*2,:)-me(:,2,i);

            dis=sqrt(xdis.^2+ydis.^2);
            b=dis<=0.01;
            c=sum(b);d=find(c~=0);

            hits(j,i,k)=sum(length(d));
            rain(i*2+1,d)=rand(1,length(d));
            rain(i*2+2,d)=ones(1,length(d));

            scatter(me(:,1,i),me(:,2,i),30,'r','filled')
            xlim([0 1])
            ylim([0 1])
            hold on
            scatter(rain(i*2+1,:),rain(i*2+2,:),'.')
            hold off

            pause(0.05)
%             L(i)=getframe(gca);
            i=i+1;
        end
    end
end
```

```matlab
    end

ax1=axes();
scatter(1:trials,mean(sum(hits,2),3)',20,'k','filled')
hold on
errorbar(1:trials,mean(sum(hits,2),3),std(sum(hits,2),1,3),'vertical','LineStyle','none','Color','k')
hold off
ylim(ax1,[0 max(mean(sum(hits,2),3))+2*max(std(sum(hits,2),1,3))])
xlabel(ax1,"Multiples of " + 0.1*v/rainspeed + "X speed of rain");
ylabel(ax1,'# of snowflake hits')
xlim(ax1,[0 trials+1])
xticks(ax1,1:trials)
ax2=axes('Position',get(ax1,'Position'),'XAxisLocation','top','Color','none');
ax2.YAxis.Visible='off';
xlabel(ax2,"Multiples of " +abs(0.1*v/wind)+ "X speed of wind")
xlim(ax2,[0 trials+1])
xticks(ax2,1:trials)
set(ax1,'position',get(ax1,'position').*[1 1 1 0.95])
set(ax2,'position',get(ax2,'position').*[1 1 1 0.95])

% video=VideoWriter('Rain running','MPEG-4');
% video.FrameRate=50;
% open(video);
% writeVideo(video,L);
% close(video)
```

## Results

Based on the average magnitude of rain terminal velocity of 10 m/s [7], wind speed of 1.3 m/s in New York City from 2005 to 2023 [8], and human walking speed of 1 m/s [9], the average number of collisions with raindrops was simulated as a function of walking and horizontal wind speed. Regardless of horizontal wind or vertical rain speed, the average number of collisions decreased with increased walking speed (Fig. 1A-D). Additionally, walking along the same x-direction as the horizontal component of the net rain vector, namely, the horizontal wind speed, resulted in fewer raindrop collisions (Fig. 1A,C) at lower walking speeds than walking along the opposite direction (Fig.1 B,D). With wind speed at 1/10X rain speed, the average number of raindrop hits rapidly decreased with increased walking speed, approaching an asymptote at 1.5-2X wind speed when walking in the same direction as the horizontal component of the net rain vector (Fig. 1A,C) and at 4-5X windspeed when walking in the opposite direction (Fig. 1B,D). Increasing the density of raindrops from 250/field (low) to 1,000/field (high) appeared to have only resulted in a proportional increase in raindrop hits.

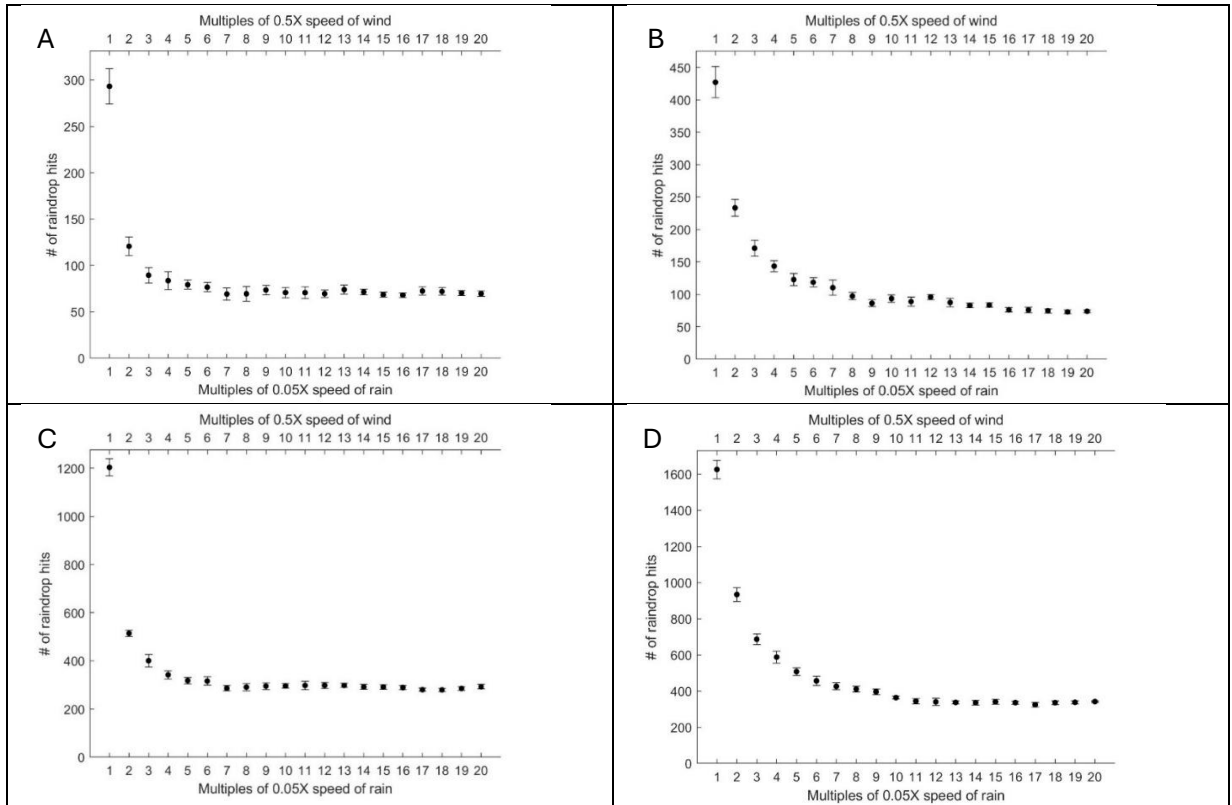

Figure 1: Simulation of wetness when running in the rain. Wind speed 1/10X of rain speed. 250 raindrops/field in (A) and (B), with x-direction of running and rain in same direction in (A) and opposite direction in (B). 1000 raindrops/field in (C) and (D), with x-direction of running and rain in same direction in (C) and opposite direction in (D). n=10 for each datapoint. Error bars=standard deviation.

With wind speed set to 1/2X rain speed, we noticed the appearance of a local minimum at 1X wind speed when walking in the same direction as the horizontal component of the net rain vector (Fig. 2A,C). Following the local minimum, the average number of raindrop hits increased slightly, reaching an asymptote at 1.5X wind speed for both low and high rain densities. No local minimum appeared when walking in the opposite direction of the horizontal component of the net rain vector (Fig. 2C,D). Similarly, an asymptote was approached at 1.5X wind speed for both rain densities. With wind speeds set to 1X and 2X rain speeds, similar local minimums appeared at 1X wind speed only when walking with the horizontal component of the rain (Fig. 3A,C, Fig. 4A,C). In all cases, an increase in the initial walking speed resulted in a rapid decrease in raindrop collisions.

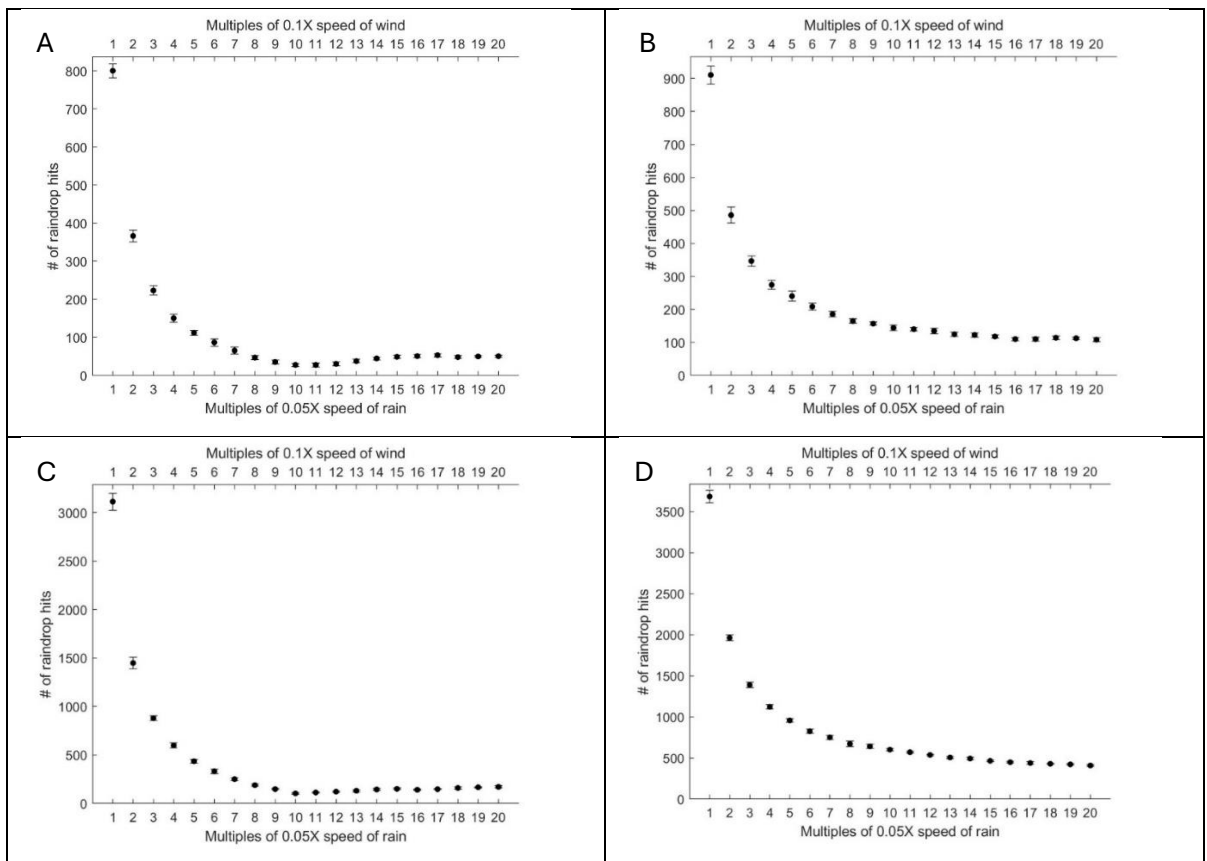

Figure 2: Simulation of wetness when running in the rain. Wind speed 1/2X of rain speed. 250 raindrops/field in (A) and (B), with x-direction of running and rain in same direction in (A) and opposite direction in (B). 1000 raindrops/field in (C) and (D), with x-direction of running and rain in same direction in (C) and opposite direction in (D). n=10 for each datapoint. Error bars=standard deviation.

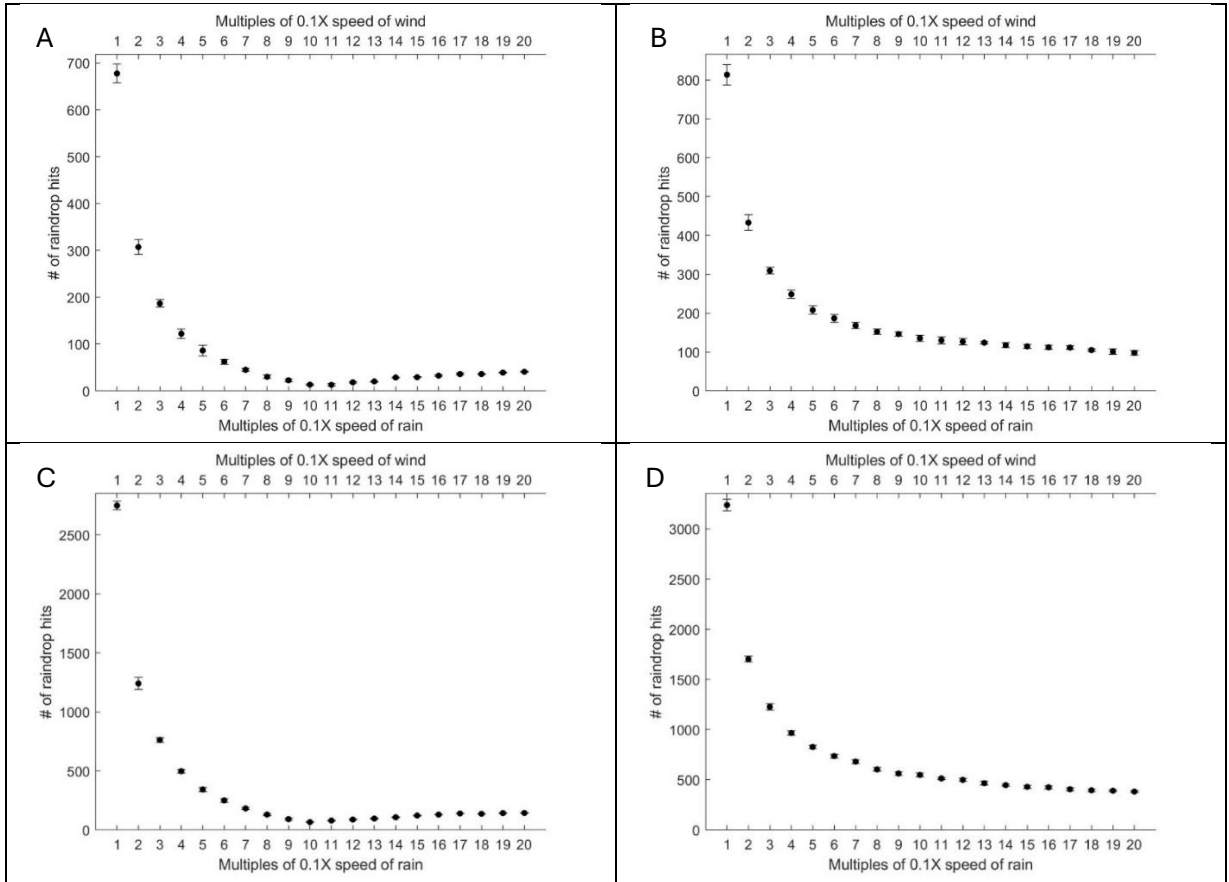

Figure 3: Simulation of wetness when running in the rain. Wind speed 1X of rain speed. 250 raindrops/field in (A) and (B), with x-direction of running and rain in same direction in (A) and opposite direction in (B). 1000 raindrops/field in (C) and (D), with x-direction of running and rain in same direction in (C) and opposite direction in (D). n=10 for each datapoint. Error bars=standard deviation.

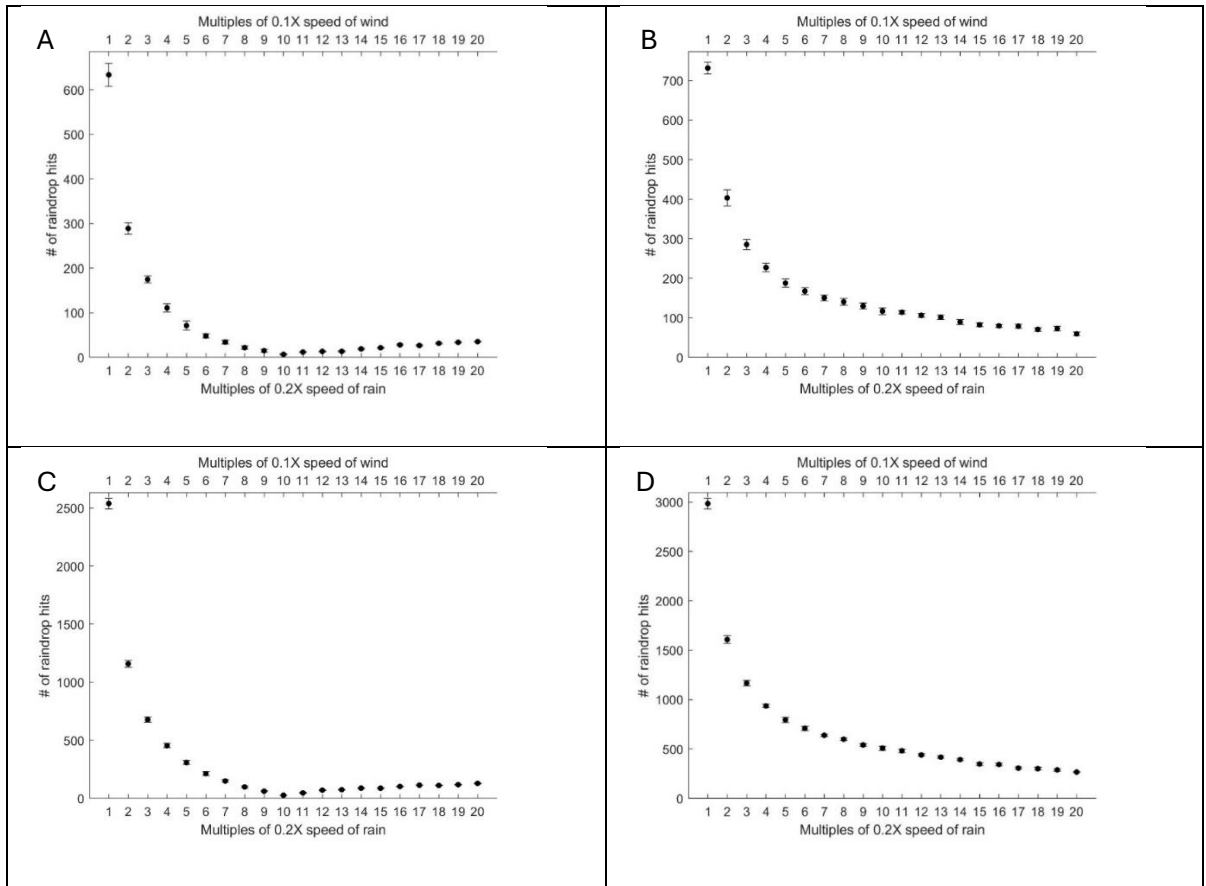

Figure 4: Simulation of wetness when running in the rain. Wind speed 2X of rain speed. 250 raindrops/field in (A) and (B), with x-direction of running and rain in same direction in (A) and opposite direction in (B). 1000 raindrops/field in (C) and (D), with x-direction of running and rain in same direction in (C) and opposite direction in (D). n=10 for each datapoint. Error bars=standard deviation.

We next adjusted the ratios between wind speed, walking speed, and rain speed to repurpose the simulation for snowfall. Based on the average magnitude of snowfall terminal velocity of 1 m/s [10], we simulated the average number of snowflake collisions as functions of wind speed and walking speed. With a wind speed of 1/10X snowfall speed, the number of snowflake hits decreased with increased walking speed (Fig. 5A,B). Walking in the same direction as the horizontal component of the net snowfall vector led to an asymptote at 3X windspeed, while walking against the horizontal component vector led to an asymptote at 12X windspeed. Additionally, walking against the wind resulted in greater snowflake collisions at lower walking speeds.

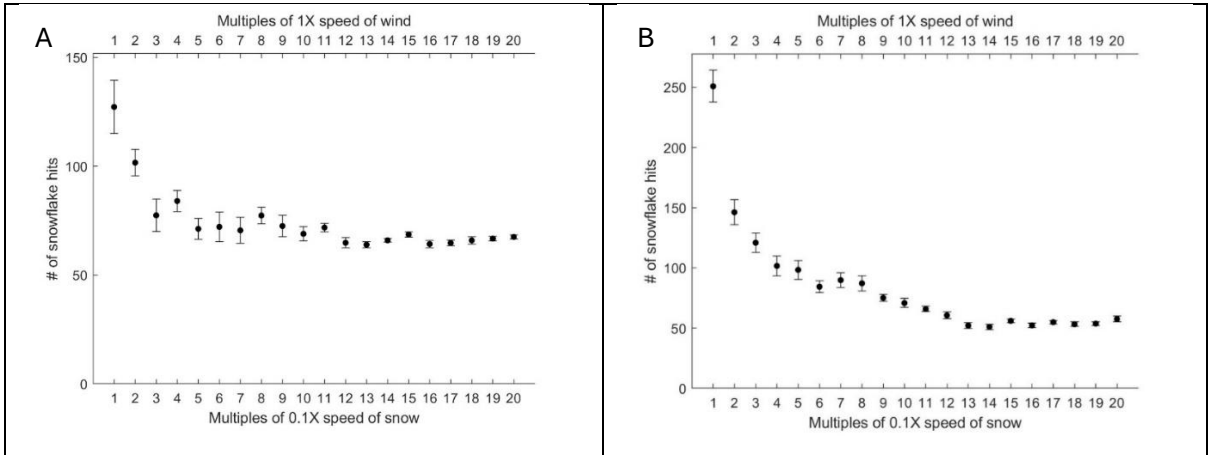

Figure 5: Simulation of wetness when running in the snow. Wind speed 1/10X of snow speed. 250 snowflakes/field in (A) and (B), with x-direction of running and rain in same direction in (A) and opposite direction in (B). n=10 for each datapoint. Error bars=standard deviation.

With wind speed adjusted to 1X snowfall speed, a local minimum appeared at 1X wind speed when walking along the horizontal component of the net snowfall vector (Fig. 6A). Afterwards, the number of snowflake hits increased ~10X at a walking speed of 4X wind speed. No local minimum was seen when walking against the wind, with an asymptote appearing at 2-3X windspeed. When wind speed was adjusted to 5X and 10X snowfall speed, similar local minimums appeared at 1X windspeed only when walking with the horizontal vector (Fig. 7A, Fig. 8A). Similar to raindrops, overall, an increase in initial walking speed resulted in a decrease in snowflake hits.

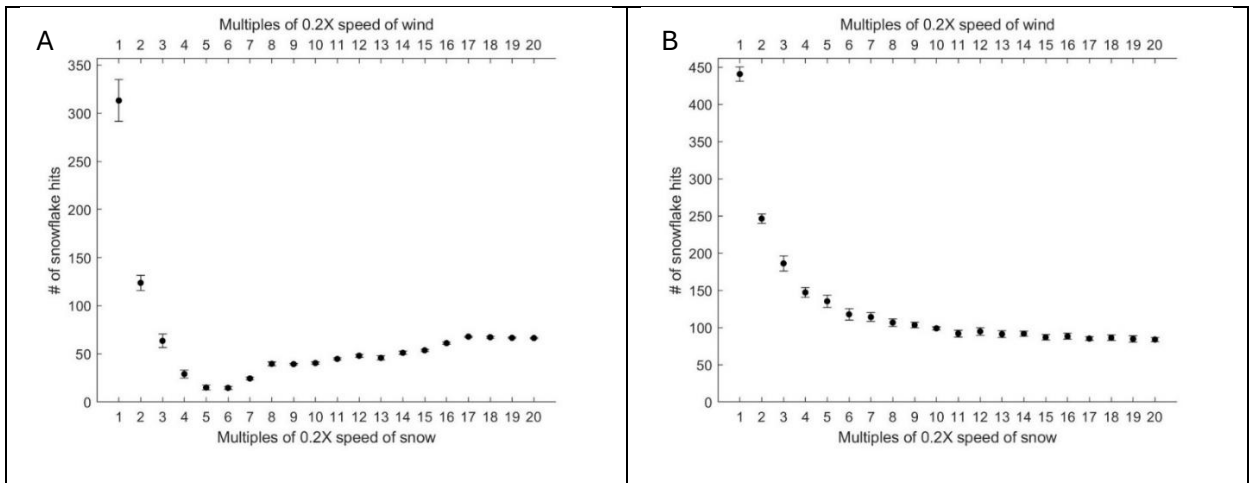

Figure 6: Simulation of wetness when running in the snow. Wind speed 1X of rain speed. 250 snowflakes/field in (A) and (B), with x-direction of running and rain in same direction in (A) and opposite direction in (B). n=10 for each datapoint. Error bars=standard deviation.

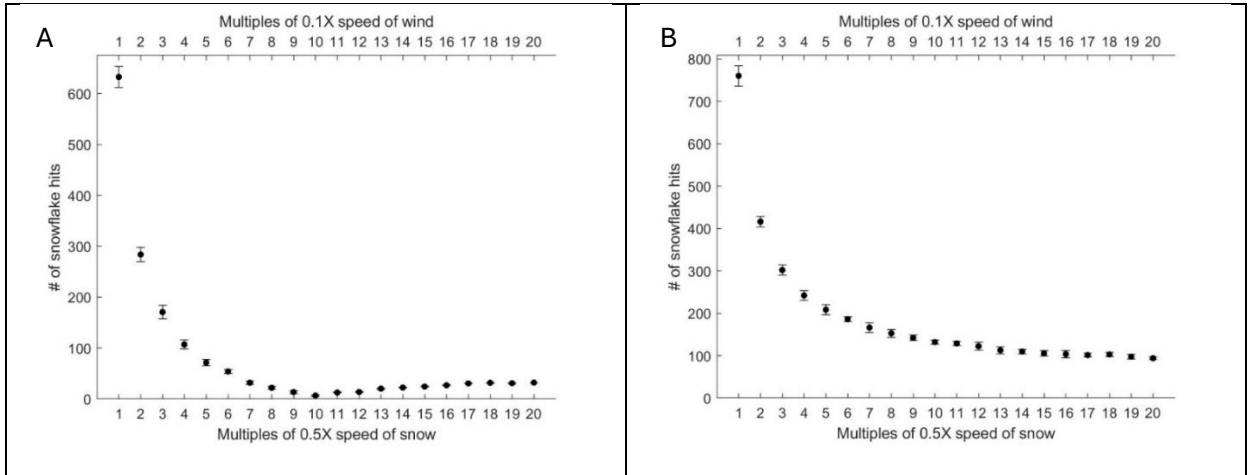

Figure 7: Simulation of wetness when running in the snow. Wind speed 5X of snow speed. 250 snowflakes/field in (A) and (B), with x-direction of running and rain in same direction in (A) and opposite direction in (B). n=10 for each datapoint. Error bars=standard deviation.

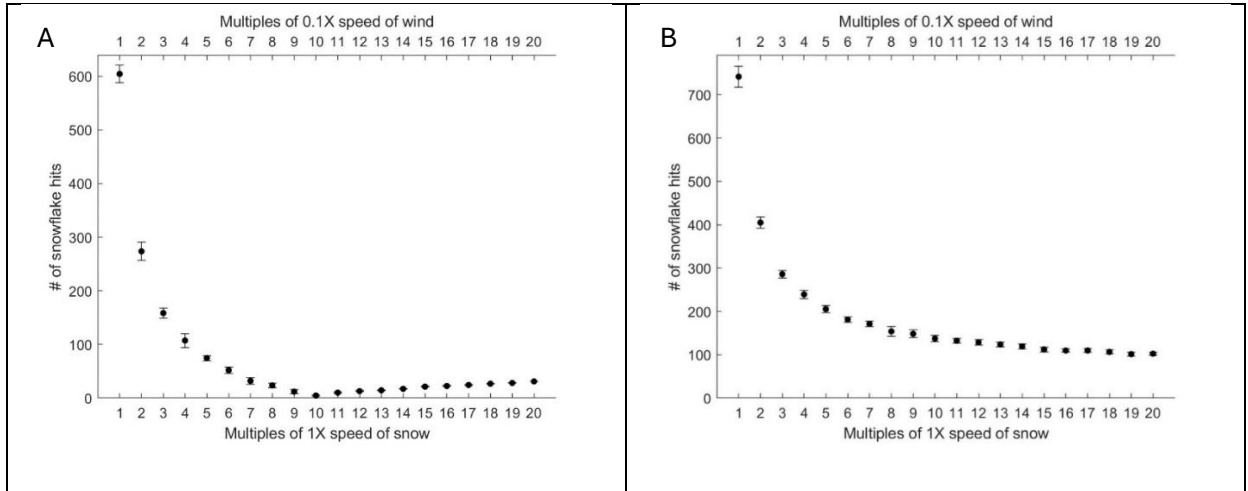

Figure 8: Simulation of wetness when running in the snow. Wind speed 10X of snow speed. 250 snowflakes/field in (A) and (B), with x-direction of running and rain in same direction in (A) and opposite direction in (B). n=10 for each datapoint. Error bars=standard deviation.

In cases of snow flurry, when there is no net wind vector and therefore no net horizontal component of the snowfall vector, an increase in the initial walking speed also resulted in a decrease in snowflake collisions, reaching an asymptote at 0.6X snowfall speed (Fig. 9A).

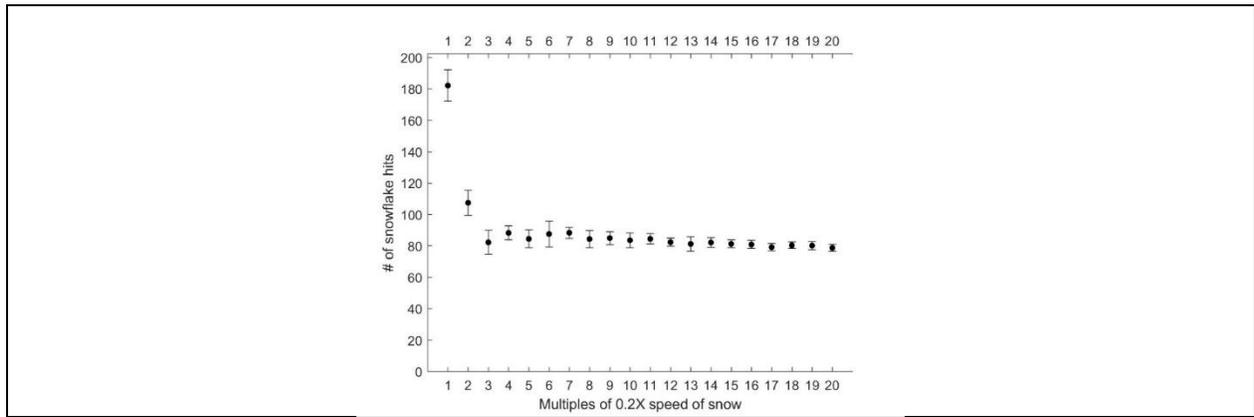
Figure 9: Simulation of wetness when running in the snow. Snow flurry, falling direction randomly between ±45° from vertical. 250 snowflakes/field. n=10 for each datapoint. Error bars=standard deviation.

**Discussion**

The question of whether one should walk or run in the rain to stay as dry as possible has been considered many times over the past few decades. Our conclusions using a MATLAB generated simulation of precipitation, that 1) moving faster initially always results in decreased wetness, 2) moving in the same direction as the horizontal component of the net rain vector, namely, the horizontal wind speed, allows one to approach the minimum number of collisions faster than in the opposite direction, and 3) a local minimum in collisions appears only when one moves in the same direction as the horizontal component of the net rain vector are in line with others who have approach this problem through different methods.

In theory, assuming no horizontal component of rainfall, the minimum number of raindrop collisions is approached as one achieves increasingly faster speeds. Moving at many orders of magnitude faster than the speed of raindrops would result in one tracing out a cylinder of rain parallel to the direction of movement, while no rain collisions are experienced on any orthogonal surfaces. Assuming a constant density of rain per unit time, this would be the global minimum amount of raindrop hits.

Our simulation was based on realistic assumptions of walking speed with respect to rain and wind speed. Under these conditions, the horizontal component of rainfall played a role in whether one can achieve a local minimum in rain drop hits. Namely, if one moved with the same horizontal direction and magnitude as the net rain vector, namely, the horizontal wind speed, an "optimum" number of raindrop collisions can be reached. By traversing horizontally at the same speed as the horizontal speed of the raindrops, one would minimize the number of collisions on surfaces whose normal vectors are parallel to the direction of motion. The only collisions would be experienced on orthogonal surfaces, the cumulative amount of which would be less than the sum of collisions along the parallel and orthogonal surfaces at slower or faster speeds.

Using our simulation, we were able to perform the same analysis with snowfall. As snowflakes have a much greater surface area than raindrops, they are subject to higher magnitudes of air resistance and thus reach a slower terminal velocity earlier. Similar to rain, moving faster initially results in fewer collisions with snowflakes. However, a key difference with rainfall is that a global minimum appears to be achievable without approaching very high speeds. The terminal speed of snowfall is comparable to normal human walking speed. Therefore, one would experience more collisions by moving at high speeds along the same direction as the horizontal component of the net snowfall vector. By moving with the same speed as the horizontal component, the number of collisions along the parallel direction of movement is minimized. While the number of collisions along the orthogonal direction would increase, the total number of collisions along the parallel and orthogonal surfaces would decrease due to the slower terminal speed of snowfall.

    In summary, one should run rather than walk in the rain or snow to stay as dry as possible. If one moves against the wind, running as fast as possible would result in colliding with the least amount of precipitation. If one moves with the wind, matching the horizontal speed of the raindrops or snowflakes would be the optimal speed to achieve the local minimum of collisions.